\documentclass[aps,prd,twocolumn,eqsecnum,amssymb,amsmath,showpacs,a4paper, superscriptaddress]{revtex4-2}

\usepackage{graphicx}
\usepackage{amsfonts}
\usepackage{amsmath}
\usepackage{hyperref}
\usepackage{units}
\usepackage{color}
\usepackage{dcolumn}
\usepackage{bm}
\usepackage{mathrsfs}
\usepackage{float}
\usepackage{cleveref}
\usepackage[normalem]{ulem}
\usepackage{appendix}
\usepackage{mathtools}
\usepackage{esvect}
\usepackage[normalem]{ulem}

\usepackage{array,multirow,graphicx}

\newcommand{\Sp}{\mathrm{sp}}
\newcommand{\tp}{\mathrm{tp}}

\newcommand{\Tloc}{\mathbf{T}}
\newcommand{\Rloc}{\mathbf{R}}

\begin{document}

\title{Imperfect draining vortex as analogue extreme compact object}
\author{Th\'eo Torres}
\email{theo.torres_vicente@kcl.ac.uk}
\affiliation{Department of Physics, King’s College London, The Strand, London WC2R 2LS, UK}

\author{Sam Patrick}
\email{sampatrick31@gmail.com}
\affiliation{Department of Physics and Astronomy, University of British Columbia, Vancouver, British Columbia, V6T 1Z1, Canada}
\affiliation{School of Mathematical Sciences, University of Nottingham, University Park, Nottingham, NG7 2RD, UK}

\author{Ruth Gregory}
\email{ruth.gregory@kcl.ac.uk}
\affiliation{Department of Physics, King’s College London, The Strand, London WC2R 2LS, UK}
\affiliation{Perimeter Institute, 31 Caroline St, Waterloo, Ontario N2L 2Y5, Canada}

\date{\today}

\begin{abstract}
    Motivated by recent experimental progress, we study scalar wave propagation over an imperfect draining vortex, which can serve as an analogue for rotating and non-rotating extreme compact objects (ECOs). 
    We encapsulate the absorbing properties of the analogue ECO by means of an effective boundary located around the analogue horizon.
    The presence of reflection at the effective boundary, characterised by a single parameter $\mathcal{K}$, allows for the existence of bound states located between the effective vortex core and the angular momentum barrier.
    The existence of these bound states leads to an enhanced absorption when the frequency of the incoming wave matches bound state frequencies, which result in Breit-Wigner type spectral lines in the absorption spectra.
    We also investigate the case of rotating analogue ECOs. In this scenario, some of the bound states undergo superradiant amplification and become unstable. 
    In both the rotating and non-rotating case, we calculate numerically transmission/reflection spectra exhibiting the enhanced absorption/amplification. 
    We complement our numerical study with WKB estimates as well as an extension of the P\"oschl-Teller toy model which we solve analytically.
    Our simple model exhibits distinctive properties which could be observed in future analogue gravity experiments. We further argue that the observation of the spectral lines could be a way to characterise the effective field theory at play in the vicinity of the vortex core.
\end{abstract}

\maketitle
\section{Introduction}

Recent years have seen an increase in efforts to experimentally observe various effects predicted in classical and quantum field theory in curved spacetimes using condensed matter platforms.
The quest to probe elusive phenomena in systems ranging from classical hydrodynamics and Bose-Einstein condensate to non-linear optics has been made possible by an analogy between excitations in a moving fluid and scalar fields propagating in a curved spacetime~\cite{Unruh:1980cg,Barcelo:2005fc}. 
This analogy has already stimulated experiments reporting on the observation of Hawking radiation~\cite{Weinfurtner:2010nu,euve2016observation,euve2020scattering,steinhauer2016observation,kolobov2021observation}, superradiance~\cite{Torres:2016iee,braidotti2022measurement}, quasinormal mode oscillations~\cite{Torres:2020tzs} and cosmological particle production~\cite{Eckel:2017uqx}.
While the original analogy offered a novel perspective as well as a new set of tools to study a multitude of fundamental effects, it nonetheless relies on strong assumptions on the system under investigation which rarely apply perfectly in experimental settings.

In this paper, we focus on the scattering of waves with a draining vortex flow, a system which can be thought of as an analogue of a rotating black hole. Two key features of rotating black holes, namely the ergoregion and horizon, are located in a vortex where the total velocity of the fluid and the radial velocity become equal to the propagation speed of the waves respectively. Due to the large flow velocities close to the vortex core, it is expected that extra hydrodynamical effects, such as vorticity~\cite{patrick2018black}, may be present and spoil the simplistic draining bathtub model of the vortex whose core is a perfect absorber, e.g. \cite{visser1998acoustic,basak2003superresonance,basak2003reflection,basak2005analog,berti2004qnm,dolan2012resonances,richartz2015rotating,churilov2019scattering}.
To this end, we introduce an effective model in the form of a draining vortex with a partially reflective boundary condition inside the core.
Hints of core reflections have already been found in existing analogue gravity experiment, e.g. in \cite{Torres:2020ckk} it was shown that the reflection coefficient for high frequency counter-rotating azimuthal modes tends to a finite constant, which is suggestive of reflection occurring in the core.

In our model, the effective boundary condition at the centre of the vortex is located a small distance outside the analogue horizon and is described by a single quantity, $\mathcal{K}$. Our model does not aim to describe the underlying phenomena responsible for the imperfect absorption at the vortex core, rather, we focus on the signature of the presence of core reflections, making our findings applicable not only to the draining bathtub vortex but to a wide class of systems (namely, ones containing a horizon and ergosphere).
We show that the presence of this effective boundary condition results in an enhanced absorption of the waves for specific frequencies, which are visible through the presence of spectral lines in the transmission spectrum. 
This effect has been studied in an astrophysical setting, where extremely compact objects (ECOs) surrounded by an effective boundary also exhibit sharp lines in the absorption cross section~\cite{Macedo:2018yoi}. The case of rotating compact objects with similar boundary conditions was studied in~\cite{Maggio:2017ivp}.

We will see that in the case of a rotating vortex, this enhanced absorption coupled with rotational superradiance causes a drastic amplification of the incoming wave at particular frequencies. 
Superradiant amplification is the result of negative wave-energy inside the ergosphere \cite{brito2020superradiance} and when this negative energy is absorbed, the escaping part of the wave is amplified and the overall reflection coefficient exceeds unity.
Due to the imperfect absorption in the core, part of the negative energy wave can become trapped between the reflective core and the angular momentum barrier, leading to instabilities.
For the flow parameters considered, we find that in the vicinity of instabilities the reflection coefficient may be four orders of magnitude greater when an effective boundary is introduced in the core.
We characterise the dependence of the reflection coefficient on the absorption parameter $\mathcal{K}$ and find that for each superradiant mode, there exists an optimal value of $\mathcal{K}$ which maximises the reflection coefficient.

The paper is organised as follows. In section~\ref{sec:model} we introduce our model and the various quantities of interest. Section \ref{sec:toymodel} presents an analytic approximation to the problem using matched P\"oschl-Teller potentials to gain insight into the likely phenomena of interest. In section~\ref{sec:spectral_lines}, we solve numerically the equations of motion for our system both for the non-rotating and rotating draining vortex. We show the presence of spectral lines in the transmission and reflection coefficient spectra. The numerical study is accompanied by theoretical predictions based on a WKB analysis in section \ref{sec:WKB}. Finally, we conclude our study with a discussion of our results in section~\ref{sec:conclusion}.

\section{Model, wave equation and effective core}\label{sec:model}

In this paper we are interested in the propagation of surface gravity waves over a flowing fluid. The fluid is considered to be inviscid and irrotational such that its velocity field can be expressed in terms of a single scalar potential $\Phi$. Waves are small perturbation, $\phi$, around the background velocity field given by $\vec{v}_0 = \nabla \Phi_0$. For shallow water waves, the equation of motion for the propagation of the perturbation $\phi$ is given by,
\begin{equation}\label{eq:fluid_eq}
    \mathcal{D}_t^2 \phi - c^2 \Delta \phi = 0,
\end{equation}
where $\mathcal{D}_t = \partial_t + \vec{v}_0\cdot\nabla$ is the material derivative, $Delta$ is the Laplacian, and $c$ is the propagation speed of the waves \cite{schutzhold2002gravity}. In the following we will consider that the background flow velocity $\vec{v}_0$ is the standard draining bathtub (DBT) flow, see e.g. \cite{dolan2012resonances}, given by,
\begin{equation}
    \vec{v}_0 = -\frac{D}{r}\vec{e}_r +  \frac{C}{r}\vec{e}_\theta, 
\end{equation}
where $(\vec{e}_r,\vec{e}_\theta)$ is the orthonormal basis in polar coordinates $(r,\theta)$.
The parameters $C$ and $D$ respectively characterise the rotation and drain of the vortex, and are taken to be constants.
To make the connection between surface waves and fields around a black hole, one recognizes that Eq.~\eqref{eq:fluid_eq} is exactly the Klein-Gordon equation for a massless scalar field in a curved spacetime with line element,
\begin{equation}
    ds^2 = -c^2dt^2 + \left(dr + \frac{D}{r}dt\right)^2 + \left(rd\theta - \frac{C}{r}dt\right)^2.
\end{equation}
This spacetime has a horizon at $r_h = D/c$ and an ergosphere at $r_e=\sqrt{C^2+D^2}/c$.
For the rest of this work, we scale lengths and times with respect to $r_h$ and $c/r_h$ respectively, which amounts to setting $c=D=1$.
The horizon is then located at $r=1$ and the ergosphere at $r=\sqrt{C^2+1}$.

Since the flow is considered stationnary and axisymmetric, we may decompose the perturbation $\phi$ onto the following mode basis
\begin{equation}
    \phi = \sum_m \int  R(r) e^{-i\omega t + i m \theta} d\omega,
\end{equation}
where $\omega$ and $m$ are respectively the frequency and azimuthal number of each mode. Note that the radial profile $R(r)$ depends on the frequency and azimuthal number but we have omitted explicit reference to this dependence to lighten the notation. It is however understood that each mode has a different radial profile.
If we further assume that $R(r) = \phi_r(r)/\sqrt{r}$ and introduce the tortoise coordinate $r_*$ via,
\begin{equation} \label{tortoise}
    \frac{dr_*}{dr} = \frac{1}{1-r^{-2}} \ \ \Rightarrow \ \ r_*(r) = r + \frac{1}{2}\log\left(\frac{r-1}{r+1}\right),
\end{equation}
then the wave equation~\eqref{eq:fluid_eq} can be cast into the form,
\begin{equation} \label{eom}
    \frac{d^2\phi_r}{dr_*^2} - V(r)\phi_r = 0,
\end{equation}
with the effective potential $V$ defined as,
\begin{equation}\label{eq:potential}
\begin{split}
    V(r) = & -\left( \omega - \frac{m C}{r^2}\right)^2 \\
    & \qquad + \left(1-\frac{1}{r^2}\right)\left( \frac{m^2 - 1/4}{r^2} + \frac{5}{4r^4}\right).
\end{split}
\end{equation}
This encodes the competing effects of the standard angular momentum barrier, which increases the potential energy required to orbit at a particular radius, and the effect of the vortex flow field, which decreases the potential energy.

\subsection{Effective core}

As can be seen from Eq.~\eqref{eq:potential}, the effective potential asymptotes a constant at both infinity and at the horizon, that is when $r_* \rightarrow \pm \infty$. In these regions, the solutions are given as superpositions of in-going and out-going plane waves,
\begin{eqnarray}
    \phi_r \sim \begin{cases} e^{\pm i\tilde\omega r_{*}}, \quad r_*\rightarrow -\infty, \\
    e^{\pm i\omega r_{*}}, \quad r_*\rightarrow \infty,
    \end{cases}
\end{eqnarray}
where $\tilde\omega=\omega-m C$.
In the case of a purely absorbing vortex, the physical boundary condition to impose at the horizon is that the solution should contain only an in-going mode $e^{-i\tilde\omega r_{*}}$. 
In this study however, we consider that extra physical processes alter the wave propagation in the vortex core such that the vortex is now not a perfect absorber. These modifications are encompassed in an effective core where part of the wave is reflected. 
We denote the location of this effective boundary in the tortoise coordinate as $r_{*0}=r_*(r^\mathrm{wall})$, where $r^\mathrm{wall}=1+\epsilon$ with $\epsilon\ll 1$ and positive, so that \eqref{tortoise} is approximately solved by $\epsilon\simeq 2e^{2(r_{*0}-1)}$.
This allows us to evaluate quantities at the effective inner boundary to leading order in $\epsilon$.
The reflection at this boundary is characterised by the parameter $\mathcal{K}$ and the physical boundary condition is modified to 
\begin{equation} \label{wall_sol}
    \phi_r (r_* \simeq r_{*0}) \sim A^\mathrm{wall}\left(e^{-i\tilde\omega r_{*}} + \mathcal{K}e^{-2i\tilde\omega r_{*0}} e^{i\tilde\omega r_{*}}\right).
\end{equation}
For simplicity, we will restrict ourselves to $\mathcal{K}\in\mathbb{R}$ with $-1<\mathcal{K}<1$~\cite{Maggio:2017ivp}. We consider this particular range of $\mathcal{K}$ since we are interested in a vortex which is (partially) absorbing.
We note that the effect of taking $\mathcal{K}\in\mathbb{C}$ is to include for the possibility of an additional phase shift at $r_{*0}$. With this convention, Neumann boundary condition is obtained for $\mathcal{K}=1$, Dirichlet boundary condition for $\mathcal{K}=-1$ and black hole boundary condition for $\mathcal{K} = 0$.

\subsection{Absorption by an imperfect vortex}
As $r_* \rightarrow \infty$, the solution can also be written as a superposition of in-going and out-going modes,
\begin{equation}\label{infinity_sol}
    \phi_r (r_* \rightarrow \infty) \sim A^\mathrm{in} e^{-i\omega r_{*}} + A^\mathrm{out}e^{i\omega r_{*}}.
\end{equation}
From the modal constants $(A^\mathrm{in},A^\mathrm{out})$, we define the reflection and transmission coefficients,
\begin{equation} \label{scat_coefs}
|\mathcal{T}|^2 = \frac{|A^\mathrm{wall}|^2}{|A^\mathrm{in}|^2}(1 - |\mathcal{K}|^2), \qquad |\mathcal{R}|^2 = \frac{|A^\mathrm{out}|^2}{|A^\mathrm{in}|^2}.
\end{equation}
From the conservation of the Wronskian, we get that the reflection and transmission coefficients are related via
\begin{equation} \label{Refl}
    |\mathcal{R}|^2 = 1 - \frac{\tilde\omega}{\omega}|\mathcal{T}|^2.
\end{equation}
From the condition above, we see that when $\tilde{\omega}<0$, then we have $|\mathcal{R}|^2>1$ (for positive frequencies and the values of $\mathcal{K}$ we consider) and the wave has extracted energy from the vortex. This is the usual rotational superradiant amplification known, for example, in black hole physics \cite{brito2020superradiance}.
Note that the reflection at the effective inner boundary does not prevent amplification, as negative energy is still transmitted down into the draining vortex core. 
However, the amount of energy being transmitted is significantly modified by core reflections and one might therefore expect the amount of amplification to be reduced. 
This is true except for frequencies close to the frequencies of trapped modes as we will see in Sec.\ref{sec:spectral_lines}.

\subsection{Spectral lines}

The physical solution satisfying the boundary conditions Eqs.~\eqref{wall_sol} and~\eqref{infinity_sol} can be decomposed onto a basis $(u_h,u_\infty)$ where $u_h$ ($u_\infty$) satisfies a purely in-going (out-going) boundary condition at the horizon (infinity).
These solutions take the asymptotic form,
\begin{eqnarray}\label{u_h}
u_h \sim \begin{cases} e^{- i\tilde\omega r_{*}},\qquad \qquad \qquad \qquad r_*\rightarrow -\infty, \\
    A_{\infty}^{-} e^{- i\omega r_{*}} + A_{\infty}^{+} e^{+ i\omega r_{*}}, \ \ \ r_*\rightarrow \infty,
    \end{cases}
\end{eqnarray}
and,
\begin{eqnarray}\label{u_inf}
u_\infty \sim \begin{cases} A_{h}^{-} e^{- i\tilde\omega r_{*}} + A_{h}^{+} e^{+ i\tilde\omega r_{*}}, \quad r_*\rightarrow -\infty, \\
    \qquad \qquad \qquad \quad e^{+ i\omega r_{*}}, \quad r_*\rightarrow \infty.
    \end{cases}
\end{eqnarray}
Expressing $\phi_r$ as a superposition of $u_h$ and $u_\infty$, we get the following relations between the modal coefficients,
\begin{align}
\tilde\omega A_h^{+} = & \ \omega A_{\infty}^{-}, \label{modal_relation1}\\
 \omega \frac{A^\mathrm{in}}{A^\mathrm{wall}} = & \ \tilde\omega\left(A_h^+ - A_h^- \mathcal{K}e^{-2i\tilde\omega r_{*0}}\right),\label{modal_relation2}\\
  \tilde\omega A^\mathrm{wall} \mathcal{K} e^{-2i\tilde\omega r_{*0}} = & \ \omega \left(A_{\infty}^{-}A^\mathrm{out} - A_{\infty}^+ A^\mathrm{in}\right).\label{modal_relation3}
\end{align}
The transmission and reflection coefficients are singular whenever $A^\mathrm{in} = 0$, that is when,
\begin{equation}\label{eq:resonance_cond}
    \frac{A_h^+}{A_h^{-}} = \mathcal{K}e^{-2i\tilde\omega r_{*0}}.
\end{equation}
This condition defines a set of complex frequencies $\{\omega_{mn}\}$, which constitute the spectrum of resonances of the imperfect vortex.
The indices denote that for each $m$-mode there will be a different set of frequencies indexed by $n$.
Physically, these are modes which become trapped inside the cavity between the effective wall outside the horizon and the angular momentum barrier.
Since these modes tunnel out of the cavity over time, the frequency has an imaginary component.

Close to $\omega_{mn}$, $A^\mathrm{in}$ approaches zero, hence we have at first order that \mbox{$A^\mathrm{in} \sim (\omega - \omega_{mn})\partial_\omega A^\mathrm{in}\big|_{\omega = \omega_{mn}} $}.
Therefore, the transmission coefficient takes the standard Breit-Wigner form for $\omega \sim \text{Re}(\omega_{mn})$,
\begin{equation} \label{BreitWigner}
    |\mathcal{T}|^2 \sim \frac{\mathcal{A}_{mn}}{(\omega - \text{Re}(\omega_{mn}))^2 + \text{Im}(\omega_{mn})^2},
\end{equation}
with $\mathcal{A}_{mn}$ given by,
\begin{equation}
    \mathcal{A}_{mn} = \frac{|A^\mathrm{wall}|^2 (1 - |\mathcal{K}|^2)}{|\partial_\omega A^\mathrm{in}|^2_{\omega = \omega_{mn}}}.
\end{equation}
Using the relations (\ref{modal_relation1},\ref{modal_relation2} \ref{modal_relation3}) as well as the condition for frequencies to be resonant, we can evaluate the denominator as,
\begin{equation}
    \partial_\omega A^\mathrm{in}\big|_{\omega = \omega_{mn}} = \frac{\tilde\omega}{\omega}A_h^+\left( 2ir_{*0} + \partial_\omega \gamma \right)\big|_{\omega=\omega_{mn}},
\end{equation}
where $\gamma = \log(A_h^+/A_h^-)$.

\section{Toy model: the double P\"oschl-Teller potential}
\label{sec:toymodel}

To gain insight into the distribution of spectral lines and more generally in the scattering of surface waves by an imperfect vortex, we introduce here a toy model which we solve exactly. Our model approximates the effective potential~\eqref{eq:potential} by a potential for which the wave equation can be solved exactly. Our approach mirrors that used in the scattering on a Schwarzschild black hole, where a parallel is drawn with the Nariai spacetime, for which the wave equation can be put in the form (see e.g. \cite{Macedo:2018yoi}),
\begin{equation}
    \left\{ \frac{d^2}{dx^2} + \omega^2 - \frac{L^2 + 1/4}{\cosh^2x} \right\} u_{L,\omega}(x) = 0.
\end{equation}
In the Nariai spacetime, the potential is of the P\"oschl-Teller form and encapsulates the main features of the Schwarzschild potential barrier. Namely, it asymptotes a constant value, $-\omega^2$, at $x\rightarrow \pm \infty$ and has a maximum at $x=0$.
This approach was developed in~\cite{Macedo:2018yoi} to study the spectral lines of non-rotating ECOs, and their results can be directly carried over for the case of non-rotating vortex.

\begin{figure} 
\centering
\includegraphics[width=\linewidth, trim = 0.3cm 0 0.8cm 0]{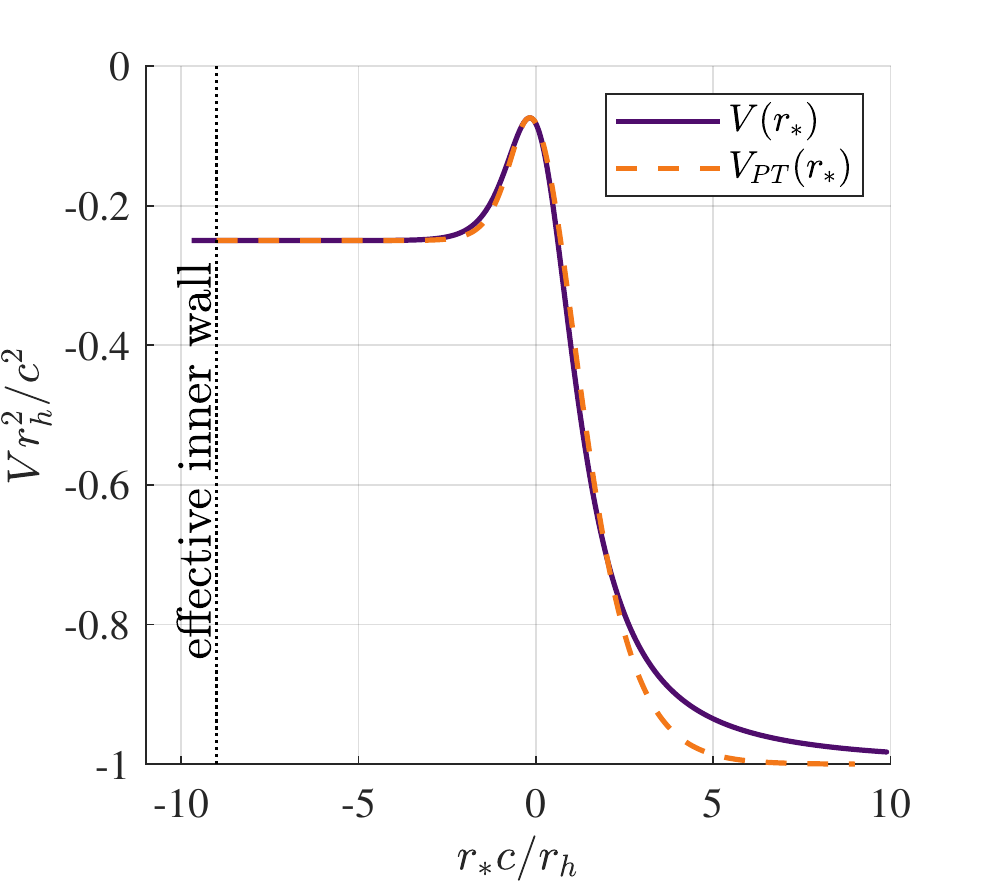}
\caption{Comparison between the true potential $V$ (purple) and the double P\"oschl-Teller approximation $V_{PT}$ (dashed red). We can see that the double P\"oschl-Teller approximation is particularly close to the true potential for $r_*<\bar{r}_*$.} \label{fig:PT_vs_trueV}
\end{figure}

When the vortex is rotating, one of the features of the potential barrier changes. The effective potential still asymptotes to $-\omega^2$ as $r_* \rightarrow \infty$ and has a single maximum but it now goes to $-\tilde{\omega}^2$ as $r_* \rightarrow -\infty$. Hence one cannot use the standard P\"oschl-Teller potential to model this problem. We can however build a new potential, $V_{PT}$, by splitting the $r_*$-axis in two regions on either side of the maximum of the potential, $\bar{r}_{*}$,
\begin{equation}
    V_{PT} = \Theta(\bar{r}_* - r_* ) V_1 + \Theta(r_* - \bar{r}_*) V_2.
\end{equation}
where $V_{1,2}$ are P\"oschl-Teller type potentials in each region, with parameters $(a_{1,2},b_{1,2},\alpha_{1,2})$, that is:
\begin{equation}
    V_{1,2}(r_*) = -\left(a_{1,2} + b_{1,2}{\mathrm{sech}}^2\left[ \alpha_{1,2}(r_* - \bar{r}_*) \right]\right).
\end{equation}
The parameters entering in the potential are chosen such that the following condition apply:
the maximum of $V_{PT}$ coincides with the one of $V$, $V_{PT}$ and $V$ have the same limit as $r_*\rightarrow \pm \infty$, $V_{PT}$ is smooth. Figure~\ref{fig:PT_vs_trueV} shows the comparison between $V_{PT}$ and the true potential. One can then solve the wave equation exactly by solving two P\"oschl-Teller like equation in each region and gluing the solutions at $r_* = \bar{r}_*$. Note that one could also avoid the splitting of radial axis by considering the more general Rosen-Morse potential~\cite{Boonserm:2010px}. However, since we are only interested in approximating the true potential $V$, this matched P\"oschl-Teller approach proves to be adequate and much simpler. We construct in Appendix~\ref{app:PT} the general solution to the wave equation with potential $V_{PT}$.

From the general solutions, one can then construct the $u_h$ and $u_\infty$ solutions which satisfy out-going boundary conditions at $\pm \infty$, and express the coefficients $(A_h^{\pm},A_\infty^{\pm})$ analytically. The exact forms of the horizon modal coefficient are given in \eqref{eq:modal_PT_hor}. One can then find the values of the resonant frequencies by plugging these expressions into Eq.~\ref{eq:resonance_cond}. The resulting condition is highly non-trivial but can be solved numerically. It is however possible to gain insight into the distribution of the spectral lines by looking at the low frequency limit of this expression.

It is possible to numerically see that \mbox{$A_{h}^{+}/A_{h}^{-} \rightarrow e^{i\sigma}$}, where $\sigma$ is some real function of $C$ satisfying \mbox{$\sigma(C\rightarrow 0) = 0$}. This implies that in the low frequency regime, the resonant frequencies are given by,
\begin{equation}\label{eq:resonance_PT}
    \omega_{mn} \sim \frac{\pi(n+1/2)}{|r_{*0}|} - \frac{\sigma}{2|r_{*0}|} + mC + i \frac{\ln{|\mathcal{K|}}}{2|r_{*0}|}.
\end{equation}
This expression reduces to the one found for the P\"oschl-Teller approximation to the non-rotating vortex~\cite{Macedo:2018yoi}.
From Eq.~\eqref{eq:resonance_PT}, we see that the spectral lines should be approximately equally spaced in frequency, with a spacing controlled by the position of the effective core. For all the cases considered here, the inner core is located at $r_{*0} = -9$, which gives a spacing of the spectral lines, $\Delta \omega = \pi/9 \approx 0.35$ which is consistent with the numerical data presented in the various figures.

\section{Spectral lines, bound states and superradiance instability}\label{sec:spectral_lines}

\subsection{Spectral lines of the non-rotating vortex}

We compute numerically the absorption cross section of our draining vortex model in the following way.
Firstly, we expand the solution of \eqref{eom} in the vicinity of the horizon using the Frobenius method and implement the desired boundary condition by imposing \eqref{wall_sol} a distance $\epsilon$ away from the horizon, i.e.\ at the effective wall.
We then take this as our initial condition and evolve \eqref{eom} out into the large $r_*$ region.
At a large value of $r_*$, we use the value of $\phi_r$ and its derivative to determine the amplitudes $(A^\mathrm{in},A^\mathrm{out})$ from \eqref{infinity_sol}, and with these values we compute the scattering coefficients according to \eqref{scat_coefs}.

\begin{figure} 
\centering
\includegraphics[width=\linewidth, trim = 0.3cm 0 0.8cm 0]{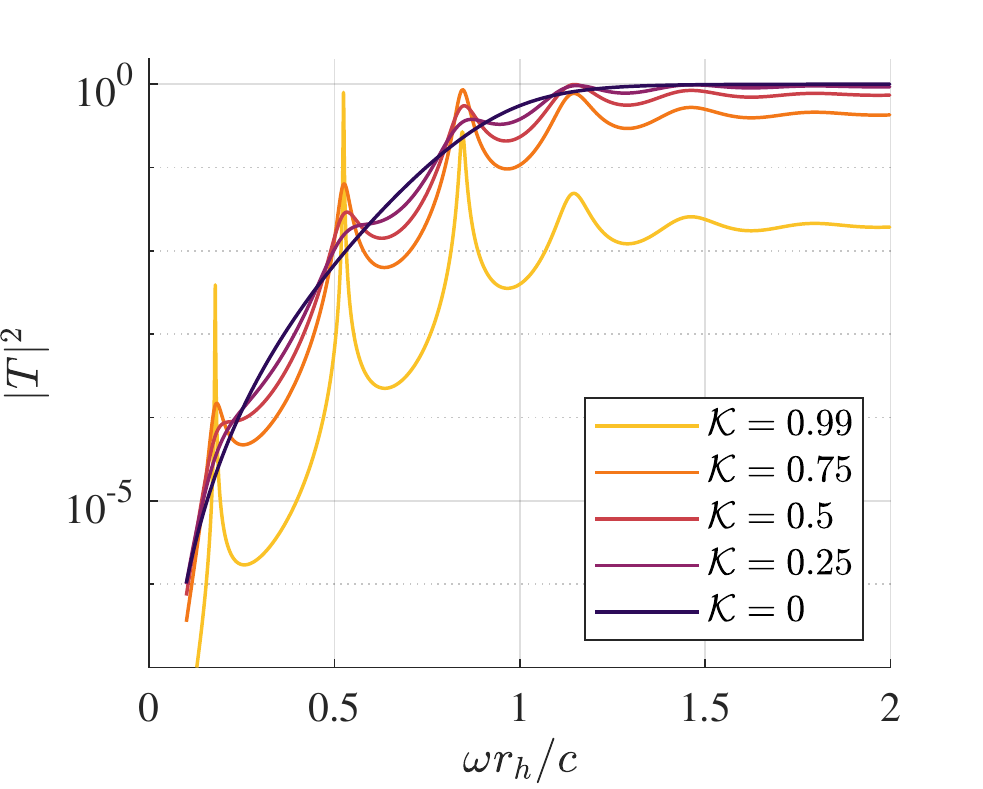}
\caption{Transmission coefficient spectrum of a non-rotating vortex flow for an azimuthal wave with $m=2$. 
The effective inner wall is located at $r_{*0} = -9$ and the amount of absorption, characterised by $\mathcal{K}$, ranges from $0$ to $0.99$. Local maxima appear at frequencies corresponding to the real part of the resonant frequencies of the system $\{\omega_{mn}\}$. These maxima become sharper as $\mathcal{K}$ increases.} \label{fig:T_spectrum_variousK}
\end{figure}

\begin{figure} 
\centering
\includegraphics[width=\linewidth,trim = 0.3cm 0 0.8cm 0]{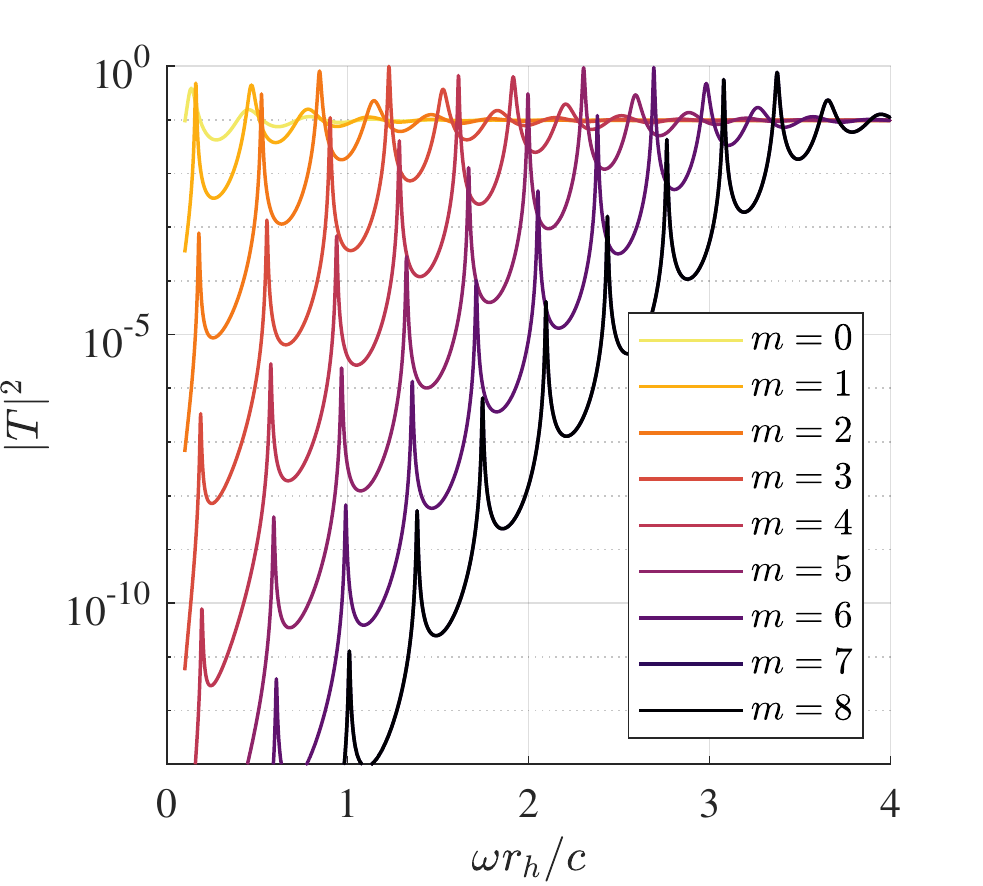}
\caption{Transmission coefficient spectrum of a non-rotating vortex flow for various azimuthal numbers $m$ ranging from $0$ to $8$. The effective inner wall is located at $r_{*0} = -9$ with $\mathcal{K}=0.95$. It is harder for the vortex to absorb higher $m$-modes due to the increased width of the angular momentum barrier.} \label{fig:T_spectrum_variousm}
\end{figure}

First, we consider the non-rotating case with $C=0$ as we expect this qualitatively to exhibit similar features to those in the ECO's~\cite{Macedo:2018yoi}.
In Fig.~\ref{fig:T_spectrum_variousK}, we display the transmission coefficient $|\mathcal{T}|^2$ as a function of $\omega$ for various values of $\mathcal{K}$.
Here, we see that absorption is enhanced in the vicinity of the \{$\omega_{mn}\}$ according to the Breit-Wigner formula \eqref{BreitWigner}
In particular, peaks of $|\mathcal{T}|^2$ are centred on $\mathrm{Re}[\omega_{mn}]$ and have a width determined by $\mathrm{Im}[\omega_{mn}]$.
Fig.~\ref{fig:T_spectrum_variousK} shows that as $\mathcal{K}$ increases, the vortex' ability to absorb a wave reduces except for frequencies close to $\omega_{mn}$.
For those frequencies, the peaks in $|\mathcal{T}|^2$ become sharper, corresponding to smaller $\mathrm{Im}[\omega_{mn}]$, i.e. longer lived resonant modes.
Fig.~\ref{fig:T_spectrum_variousm} shows the same for various $m$-modes at fixed $\mathcal{K}$.
As $m$ increases, the transmission coefficient decreases for a given frequency (unless it is near a resonance) since the size angular momentum barrier grows with $m$, making it more difficult for those $m$-modes to penetrate the vortex core.

\subsection{Spectral lines of the rotating vortex}

\begin{figure} 
\centering
\includegraphics[width=\linewidth,trim = 0.3cm 0 0.8cm 0]{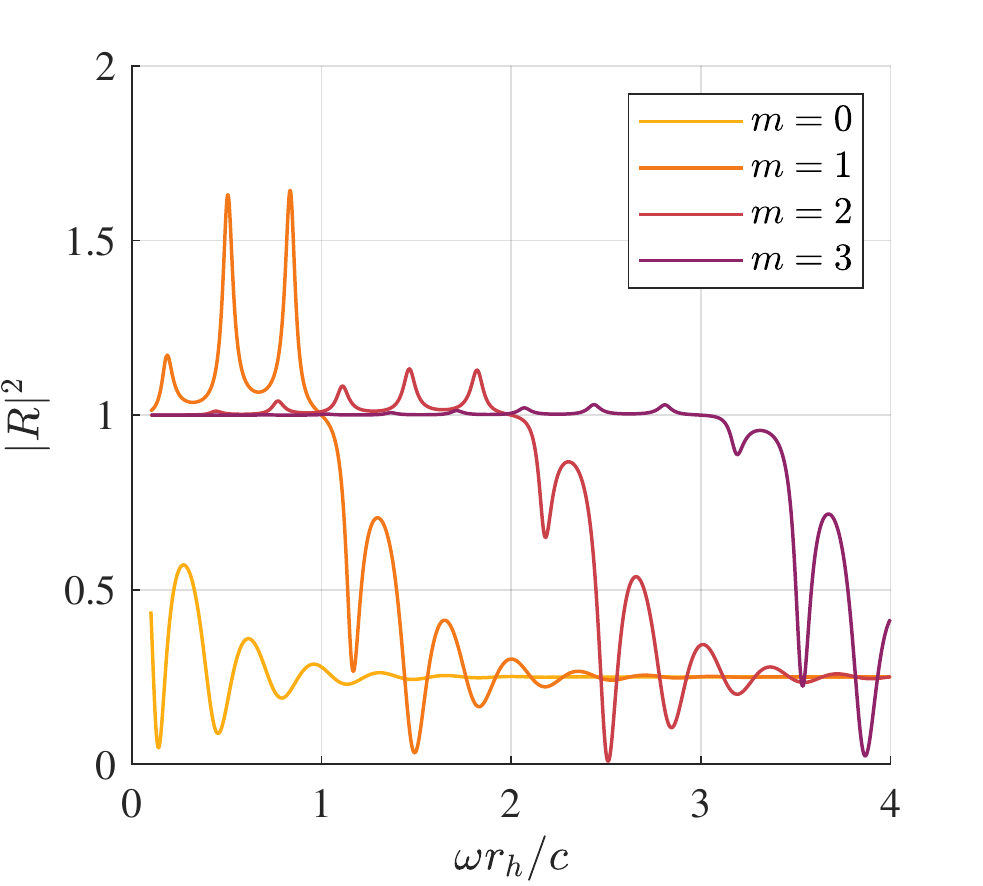}
\caption{Reflection coefficient spectra of various azimuthal waves, $0\leq m \leq 3$ incident on a rotating and draining vortex flow, with $C=D=1$ and $\mathcal{K}=0.5$. The effective inner wall is located at $r_{*0} = -9$. } \label{fig:R_spectrum_variousm}
\end{figure}

\begin{figure} 
\centering
\includegraphics[width=\linewidth,trim = 0.3cm 0 0.8cm 0]{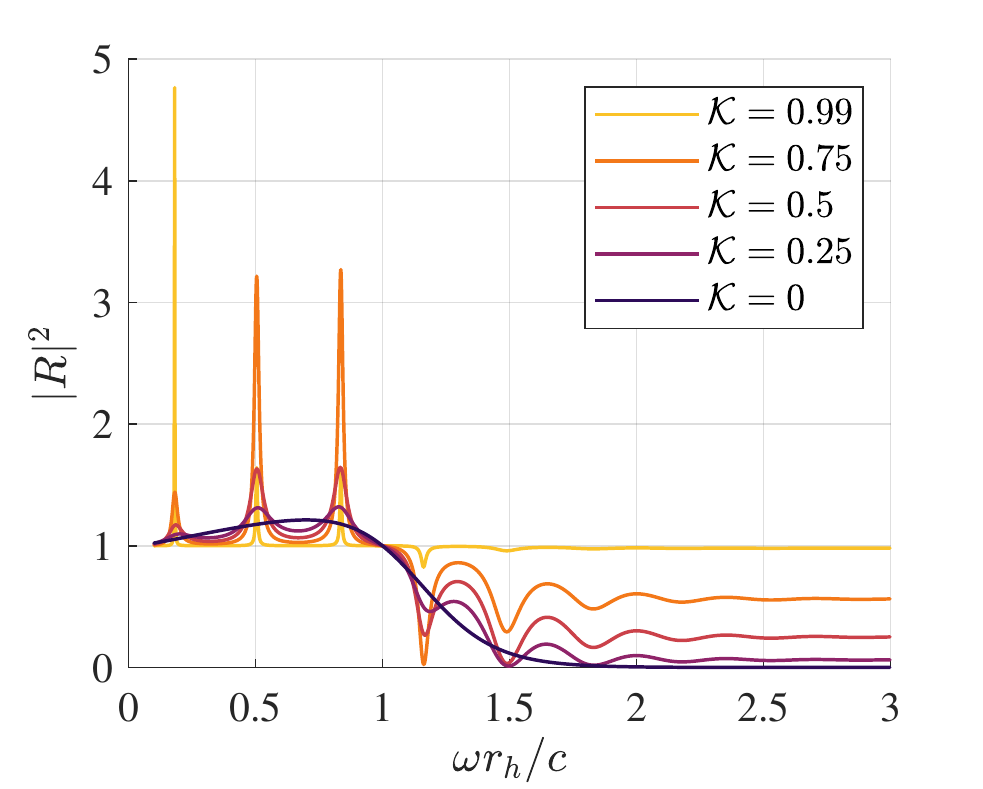}
\caption{Reflection coefficient spectrum of an $m=1$ azimuthal wave incident on a rotating and draining vortex flow, with $C=D=1$ for various values of $\mathcal{K}$ . The effective inner wall is located at $r_{*0} = -9$. } \label{fig:R_spectrum_variousK}
\end{figure}

\begin{figure} 
\centering
\includegraphics[width=\linewidth,trim = 0.3cm 0 0.8cm 0]{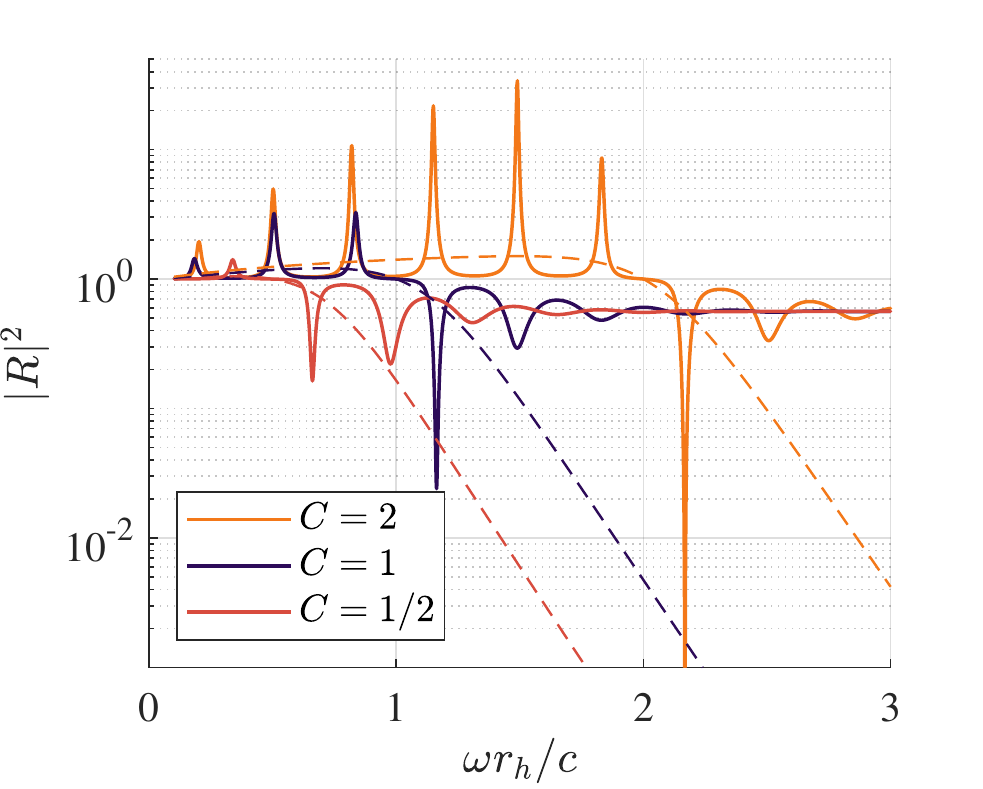}
\caption{Reflection coefficient spectrum of an $m=1$ azimuthal wave incident on a rotating and draining vortex flow for various values of the circulation parameter, 
with $\mathcal{K} = 0.75$ ($\mathcal{K} = 0$) in solid (dashed) lines. The effective inner wall is located at $r_{*0} = -9$. } \label{fig:R_spectrum_variousC}
\end{figure}

Using the same method as the previous section, we now compute the spectrum for the rotating case with $C\neq 0$.
The form of the transmission coefficient is qualitatively similar to the case of the non-rotating case, namely, it contains a series of peaks around centred on the resonant frequencies of the system.
However, when $C\neq 0$, a mode which is superradiant will have $\tilde\omega<0$. Hence, for these modes, we see from \eqref{Refl} that a peak in $|\mathcal{T}|^2$ will create peak, rather than a trough, in $|\mathcal{R}|^2$.
This means that a superradiant mode will be drastically amplified the vicinity of $\{\omega_{mn}\}$.
We display $|\mathcal{R}|^2$ at fixed $\mathcal{K}=0.5$ for various $m>0$ in Fig.~\ref{fig:R_spectrum_variousm} and as a function of $\omega$ for various $\mathcal{K}$ in Fig.~\ref{fig:R_spectrum_variousK}.
These illustrate the drastic amplification of modes with frequencies close to $\omega_{mn}$ below the superradiant threshold, and exhibit an overall similar behaviour than the transmission spectra for non rotating vortices.
We also display the effect of varying $C$ in Fig.~\ref{fig:R_spectrum_variousC} which shows that as $C$ increases, the peaks in $|\mathcal{R}|^2$ are shifted but the spacing between peaks does not significantly vary (as is expected from \eqref{eq:resonance_PT}). Furthermore, the range over which we observe drastic amplification increases (in line with the expectation that superradiance occurs for $\tilde{\omega}<0$) and the maximum value of the reflection spectrum increases with $C$ as well.

Insight can be gained by considering the energy fluxes.
Using the boundary condition \eqref{wall_sol}, we find the energy flux through the effective wall is given by,
\begin{equation}
    \mathcal{E}^\mathrm{wall} = \tilde\omega |A^\mathrm{wall}|^2(1-|\mathcal{K}|^2),
\end{equation}
whereas at infinity, the amount of energy radiated is given by,
\begin{equation}
    \mathcal{E}_\infty = \omega |A^\mathrm{out}|^2.
\end{equation}
We see that when $\tilde\omega<0$, that is when the wave satisfies the superradiance condition $\omega < mC$, then the energy flux going through the wall is negative. Hence, amplification of the reflected wave is a result of the vortex absorbing a negative energy.

\begin{figure} 
\centering
\includegraphics[width=\linewidth,trim = 0.3cm 0 0.8cm 0]{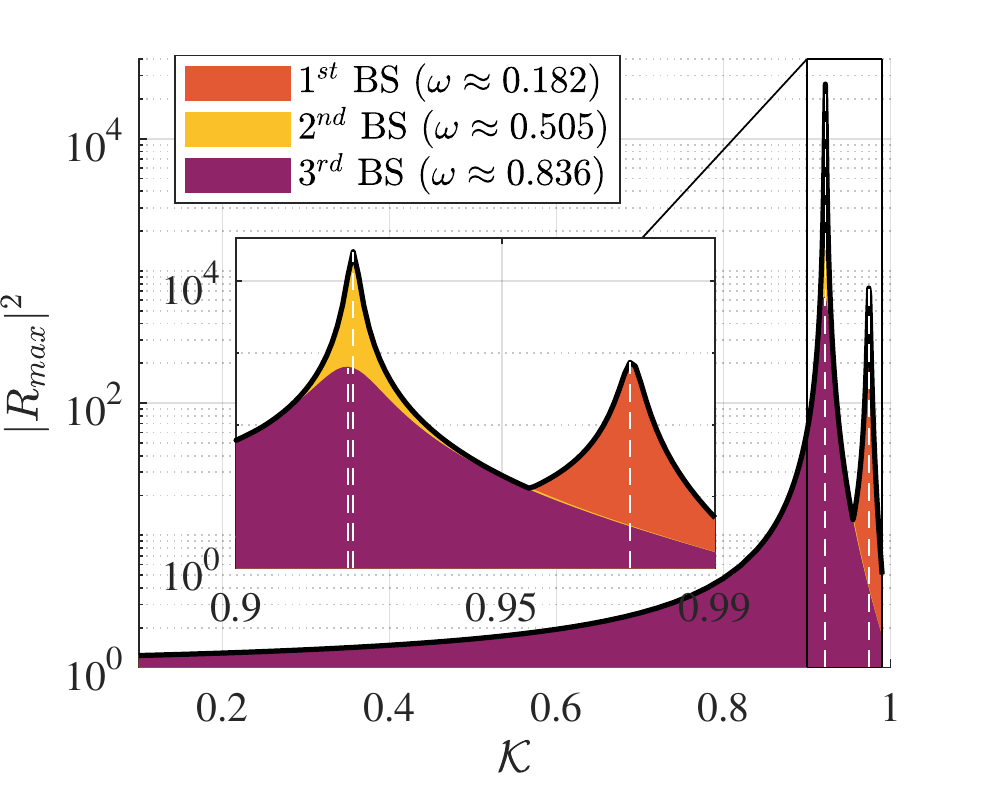}
\caption{Reflection coefficient for the 3 superradiant bound states as a function of $\mathcal{K}$. The bound states frequencies are $\omega \approx 0.182,\ 0.505$, and $0.836$, see Fig.~\ref{fig:R_spectrum_variousK}. The effective inner wall is located at $r_{*0} = -9$. We see that for each bound state frequency, the reflection coefficient reaches a maximum for different value of $\mathcal{K}$. At large $\mathcal{K}$, the first bound state ($\omega=0.182$) dominates, as $\mathcal{K}$ decreases, the second ($\omega = 0.505$) and then third ($\omega=0.836$) bound state dominates. The vertical white dashed lines, represent the value of $\mathcal{K}$ which extremises the ratio of the energy fluxes through the wall over the energy flux at infinity.} \label{fig:ref_max_3peaks_vs_K}
\end{figure}

When $\mathcal{K}=0$, the amount of energy abosrbed by the vortex is maximal. 
However, this does not mean that the amount of superradiant amplification is also maximal. 
When $0<\mathcal{K}<1$, bound states are allowed to exist between the effective wall and the potential barrier, these can then be be superradiantly amplified, extracting energy from the vortex at every cycle, leading to a superradiant instability similar to that of rotating ECOs~\cite{Maggio:2017ivp} or of a massive field around a Kerr black hole~\cite{Dolan:2007mj}.
The higher $\mathcal{K}$ is, the quicker those superradiant bound states will grow; on the other hand, a larger $\mathcal{K}$ implies a smaller energy flux through the wall.
Indeed, when $\mathcal{K}=1$, the entire energy is reflected at the wall, the flux becomes null, thus preventing superradiant amplification from occurring and resulting in a perfect reflection with $|\mathcal{R}|=1$. 
Therefore, there exists an optimal $\mathcal{K}=\mathcal{K}_\mathrm{max}$ which balances the growth of the unstable bound state and the energy flux going through the wall, resulting in a maximal reflection coefficient. 
This occurs when the ratio of the flux radiated to the flux absorbed is maximised,
\begin{equation}
    \frac{\mathcal{E}_\infty}{\mathcal{E}^\mathrm{wall}} = \frac{\omega |A^\mathrm{out}|^2}{\tilde\omega |A^\mathrm{wall}|^2 (1 - |\mathcal{K}|^2)} =  \frac{|\mathcal{R}|^2}{1-|\mathcal{R}|^2},
\end{equation}
which can be seen by taking the $\mathcal{K}$ derivative of this expression and showing that it is equal to $\partial_\mathcal{K}|\mathcal{R}|$ multiplied by a positive number.
In Fig.~\ref{fig:ref_max_3peaks_vs_K}, we show how the $n$ peaks in the reflection coefficient, i.e.\ those satisfying $\partial_\omega|\mathcal{R}|=0$, vary as a function of $\mathcal{K}$, finding that each peak reaches a maximum at a particular value of $\mathcal{K}_\mathrm{max}(\omega_{mn})$.
In the next section, we compute this value using a WKB approximation.

\section{WKB approximation}
\label{sec:WKB}

To gain further insight into the features of our system, we invoke a WKB approximation.
The first step involves applying the WKB ansatz $\phi_r(r_*) = A(r_*)e^{i\int p(r_*) dr_*}$ to the mode equation \eqref{eom}.
Assuming $|p'|\ll|p^2|$ and $|A'|\ll|pA|$, the leading and next-to-leading equations are,
\begin{equation}
    p^2+V(r_*) = 0, \qquad A(r_*)\sim\alpha|p|^{-\frac{1}{2}},
\end{equation}
where $\alpha$ is adiabatically conserved.
Since the first equation is quadratic in $p$, there will be two solutions, one of which will be radially in-going and the other out-going.
Hence at a given point $r_{*j}$ in the flow, we may write the solution,
\begin{equation}
    \phi(r_{*j}) = |p_j|^{-\frac{1}{2}}\left(\alpha_j^\mathrm{in}e^{i\int p^\mathrm{in}_j dr_*} + \alpha^\mathrm{out}_j e^{i\int p^\mathrm{out}_j dr_*} \right),
\end{equation}
noting that $|p_j^\mathrm{in}|=|p_j^\mathrm{out}|\equiv|p_j|$.

\begin{figure} 
\centering
\includegraphics[width=\linewidth]{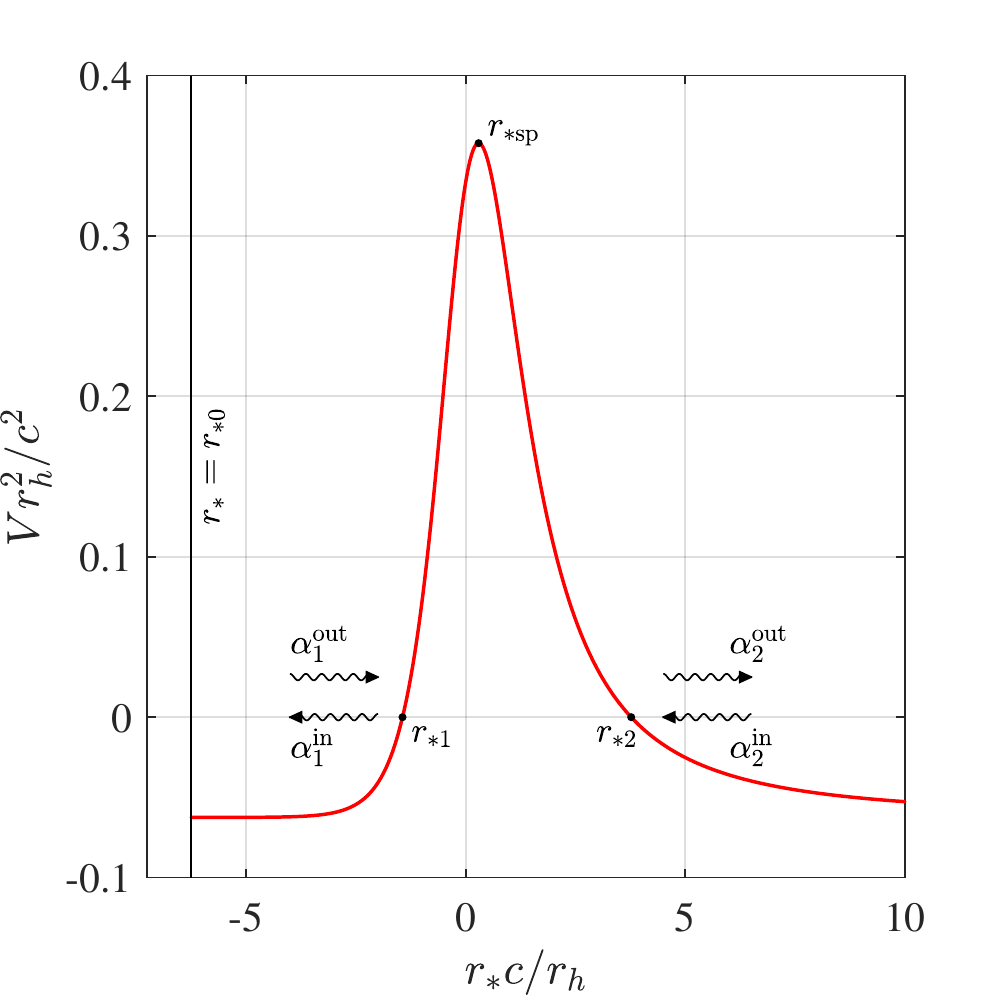}
\caption{The effective potential barrier $V$ (red) for $C=1/2$, $m=1$ and $\omega=1/4$. The effective boundary condition is placed just outside the horizon at $r=1+\epsilon$ with $\epsilon=10^{-6}$. The turning points and saddle point are illustrated as black points and the in- and out-going modes either side of the barrier also indicated.}
\label{fig:potential}
\end{figure}

We can write write an effective Hamiltonian for the system,
\begin{equation}
    H = p^2 + V,
\end{equation}
from which we obtain the ray equations,
\begin{equation}
    \dot{r}_* = \partial_p H, \qquad \dot{p} = -\partial_{r_*}H,
\end{equation}
where the overdot signifies the derivative with respect to a parameter along the rays.
The WKB approximation breaks down in the vicinity of turning points, which are the points along the rays where $\dot{r}_*=0$. From Hamilton's equations, we see that such points obey $H_{\tp}=\partial_pH_{\tp}=0$ where the latter condition immediately implies $p=0$ (which coincides with the zeros of $V$) and a divergent amplitude $A$ at these points, leading to the breakdown of the approximation there.
In that case, one can apply the connection formulae to relate the different $\alpha$ either side of the turning point, see e.g.\ \cite{patrick2020superradiance}.
If the two turning points $r_{*1}$ and $r_{*2}$ are well separated, the WKB approximation holds between them.
Then, the relation between the $\alpha$ just to the left of $r_{*1}$ and the right of $r_{*2}$ can be expressed with the transfer matrix,
\begin{equation} \label{transfer}
    \begin{pmatrix}
    \alpha^\mathrm{out}_1 \\ \alpha^\mathrm{in}_1
    \end{pmatrix} = \varepsilon\begin{pmatrix}
    \nicefrac{1}{\Tloc} & \nicefrac{-\Rloc}{\Tloc} \\
    \nicefrac{-\Rloc^*}{\Tloc^*} & \nicefrac{1}{\Tloc^*}
    \end{pmatrix} \begin{pmatrix}
    \alpha^\mathrm{out}_2 \\ \alpha^\mathrm{in}_2
    \end{pmatrix}, \quad \varepsilon = \mathrm{sgn}\left(\frac{\tilde{\omega}}{\omega}\right),
\end{equation}
where $\Rloc$ and $\Tloc$ are the local reflection and transmission coefficients respectively for a wave which scatters from the right (i.e.\ from large $r_*$) and are given by,
\begin{equation}
\begin{split}
    \Rloc^{(\tp)}_{\varepsilon=1} = & \ -i\,\frac{1-e^{-2S_{12}}/4}{1+e^{-2S_{12}}/4}, \\
    \Tloc^{(\tp)}_{\varepsilon=1} =  & \ \frac{e^{-2S_{12}}}{1+e^{-2S_{12}}/4}, \\
     \Rloc^{(\tp)}_{\varepsilon=-1} =  & \ -i\,\frac{1+e^{-2S_{12}}/4}{1-e^{-2S_{12}}/4}, \\
     \Tloc^{(\tp)}_{\varepsilon=-1} =  & \ -i\,\frac{e^{-2S_{12}}}{1-e^{-2S_{12}}/4},
\end{split}
\end{equation}
see Fig.~\ref{fig:potential} for reference.
The WKB phase integral is defined,
\begin{equation}
    S_{ij} = \int^{r_j}_{r_i}|p(r_*)|dr_*.
\end{equation}
When $r_{*1}$ and $r_{*2}$ are close to each other, the WKB approximation cannot be applied between them.
In that case, one can expand the wave equation in vicinity of a saddle point $r_{*\Sp}$ which satisfies $\partial_{r_*}H_{\Sp}=\partial_pH_{\Sp}=0$ and use the exact solution to relate the $\alpha$ on either side of $r_{*\Sp}$, see e.g.\ \cite{Torres:2020ckk}.
The result can be then expressed with the transfer matrix \eqref{transfer} but with the following expressions for the local scattering coefficients,
\begin{equation}
\begin{split}
    \Rloc^{(\Sp)}_{\varepsilon=1} = & \ -i\frac{2^{ib}}{\sqrt{2\pi}}\Gamma(\tfrac{1}{2}+ib), \\
    \Tloc^{(\Sp)}_{\varepsilon=1} =  & \ e^{-\pi b}\frac{2^{ib}}{\sqrt{2\pi}}\Gamma(\tfrac{1}{2}+ib), \\
    \Rloc^{(\Sp)}_{\varepsilon=-1} =  & \ \frac{2^{-ib}\sqrt{2\pi}}{\Gamma(\tfrac{1}{2}-ib)}, \\
    \Tloc^{(\Sp)}_{\varepsilon=-1} =  & \ -ie^{-\pi b},
\end{split}
\end{equation}
where the parameter $b$ is given by,
\begin{equation}
    b = \frac{H}{\sqrt{-\partial_p^2H\partial_{r_*}^2H}}\Bigg|_{\Sp}.
\end{equation}
In all cases, these coefficients obey,
\begin{equation}
    |\Rloc|^2 + \varepsilon |\Tloc|^2 = 1.
\end{equation}
Note that the conventional transmission coefficient, say $\Tloc'$, which is the ratio of the $A$ rather than the $\alpha$, is related by $\Tloc' = |p_0/p_\infty|^\frac{1}{2}\Tloc$, which gives the usual scattering coefficient relation in BH physics, e.g.\ \cite{brito2020superradiance}.

\begin{figure} 
\centering
\includegraphics[width=\linewidth,trim = 0.3cm 0 0.8cm 0]{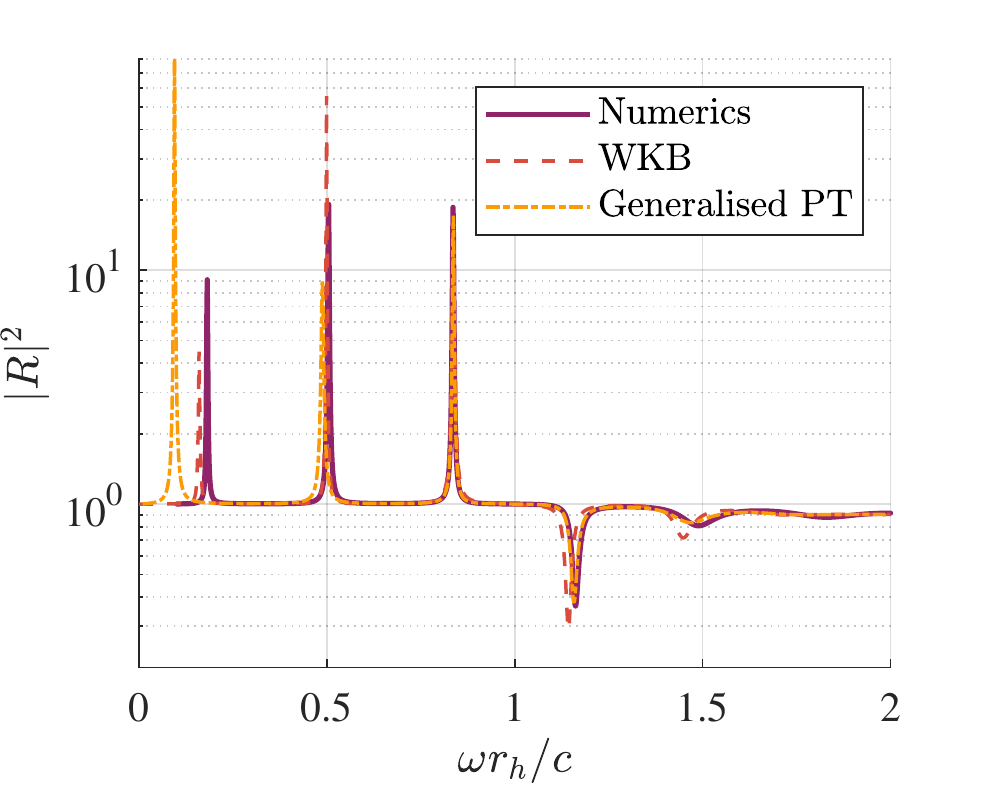}
\caption{Numerical and analytical reflection coefficient  for the $m=1$ mode incident on a vortex flow with $C=1$ with an inner core located at $r_{*0} = -9$ and with $\mathcal{K}=0.95$. The purple curve depicts the transmission coefficient obtained numerically while the dashed red and dotted dashed yellow curve show the WKB and generalised P\"oschl-Teller approximation respectively. Both analytical predictions agree well with the numerics and capture the esssential features.} \label{fig:R_spectrum_comparison}
\end{figure}

To find the total reflection coefficient, we can translate the WKB solution from $r_{*0}$ out to infinity, applying the transfer matrix in \eqref{transfer} to capture scattering with the effective potential barrier $V$.
The computation can be expressed succinctly in the following matrix equation,
\begin{equation}
\begin{split}
    \begin{pmatrix}
    \alpha^\mathrm{out}_0 \\ \alpha^\mathrm{in}_0
    \end{pmatrix} = \varepsilon & \begin{pmatrix}
    e^{-i\varepsilon S_{01}} & 0 \\ 0 & e^{i\varepsilon S_{01}}
    \end{pmatrix} \begin{pmatrix}
    \nicefrac{1}{\Tloc} & \nicefrac{-\Rloc}{\Tloc} \\
    \nicefrac{-\Rloc^*}{\Tloc^*} & \nicefrac{1}{\Tloc^*} 
    \end{pmatrix} \\
    & \qquad\times \begin{pmatrix}
    e^{-iS_{2\infty}} & 0 \\ 0 & e^{iS_{2\infty}}
    \end{pmatrix} \begin{pmatrix}
    \alpha^\mathrm{out}_\infty \\ \alpha^\mathrm{in}_\infty
    \end{pmatrix}.
\end{split}
\end{equation}
To find the total reflection coefficient at infinity $\mathcal{R}=\alpha^\mathrm{out}_\infty/\alpha^\mathrm{in}_\infty$, we then apply the boundary condition $\mathcal{K}=\alpha^\mathrm{out}_0/\alpha^\mathrm{in}_0$.
Solving the equation above, we find,
\begin{equation} \label{Rtot}
    \mathcal{R} = e^{2iS_{2\infty}} \frac{\Tloc\mathcal{K}e^{2i\varepsilon S_{01}}+\Tloc^*\Rloc}{\Tloc^*+\Tloc\Rloc^*\mathcal{K}e^{2i\varepsilon S_{01}}}.
\end{equation}
This expression can be compared with the numerical results in Fig.~\ref{fig:R_spectrum_comparison}.
We see that the WKB method gives good qualitative predictions for the total reflection coefficient.
In particular, the method captures the correct number of peaks (i.e.\ bound states) and gives a good estimate of their location in $\omega$.
Consequently, we can use the expression in \eqref{Rtot} to interpret our results.

The WKB method can provide insight into the unstable frequencies of the system.
The resonant frequencies of the system occur when the denominator of \eqref{Rtot} vanishes, which occurs for complex values of $\omega$.
An instability occurs when superradiant amplification overpowers the absorption at $r_{*0}$, which can only occur for $\varepsilon=-1$.
Since this is a low frequency phenomenon, the turning points $r_{*1}$ and $r_{*2}$ will be relatively far apart and we can apply the turning point approximation of the local scattering coefficients.
The denominator of \eqref{Rtot} then vanishes when the following condition is satisfied,
\begin{equation}
    e^{2i(S_{01}+\pi/4)} + |\Rloc\mathcal{K}| = 0.
\end{equation}
To solve for the complex frequency $\omega_{mn}=\omega_0+i\Gamma$, we anticipate that any instabilities will be slowly growing, i.e.\ $|\Gamma|\ll|\omega_0|$, and expand the phase integrals to split the condition into two parts,
\begin{equation} \label{WKB_instab_cond}
    \cos\left(S_{01}(\omega_0) + \frac{\pi}{4}\right) = 0, \qquad \Gamma = \frac{\log|\Rloc\mathcal{K}|}{2|\partial_\omega S_{01}|}\Bigg|_{\omega_0}.
\end{equation}
In order to have an unstable mode, we must have $|\Rloc\mathcal{K}|>1$, i.e.\ the amount of amplification due to superradiance must exceed the amount of absorption near the horizon.
Since $\mathrm{max}\{|\Rloc|\}$ monotonically decreases with increasing $|m|$, this immediately implies that there can be no unstable modes above a critical $m$.
Furthermore, this mode must fit inside the cavity between $r_{*0}$ and $r_{*1}$.

The WKB method can be used to give some perspective on the results in Fig.~\ref{fig:ref_max_3peaks_vs_K}. Since this figure is concerned with the maximum reflection coefficient for the superradiant bound states, we can again make use of the WKB expression for $\mathcal{R}$ in the turning point approximation with $\varepsilon=-1$.
Its square modulus is,
\begin{equation}
    |\mathcal{R}|^2 = \frac{\mathcal{K}^2-2|\Rloc|\mathcal{K}\sin 2S_{01}+|\Rloc|^2}{\mathcal{K}^2|\Rloc|^2-2|\Rloc|\mathcal{K}\sin 2S_{01}+1}.
\end{equation}
This expression has a maximum for,
\begin{equation}
    \mathcal{K}_\mathrm{max} = \frac{1+|\Rloc|^2+\sqrt{(1+|\Rloc|^2)^2-4|\Rloc|^2\sin^22S_{01}}}{2|\Rloc|\sin 2S_{01}}.
\end{equation}
Using \eqref{WKB_instab_cond}, we see that a superradiant bound state has $\sin 2S_{01}=1$, which implies $\mathcal{K}_\mathrm{max}= |\Rloc|^{-1}$ for these modes. Again using \eqref{WKB_instab_cond}, we find that these modes have $\Gamma=0$, i.e.\ they are stable bound states. In other words, the reflection coefficient is maximised when the amount of amplification at the angular momentum barrier is balanced by the absorption that occurs at the effective inner wall.
This makes further sense in light of the Breit-Wigner formula in \eqref{BreitWigner}, which says that $|\mathcal{T}|^2$ increases with the lifetime of the resonance.

\begin{figure} 
\centering
\includegraphics[width=\linewidth,trim = 0.3cm 0 0.8cm 0]{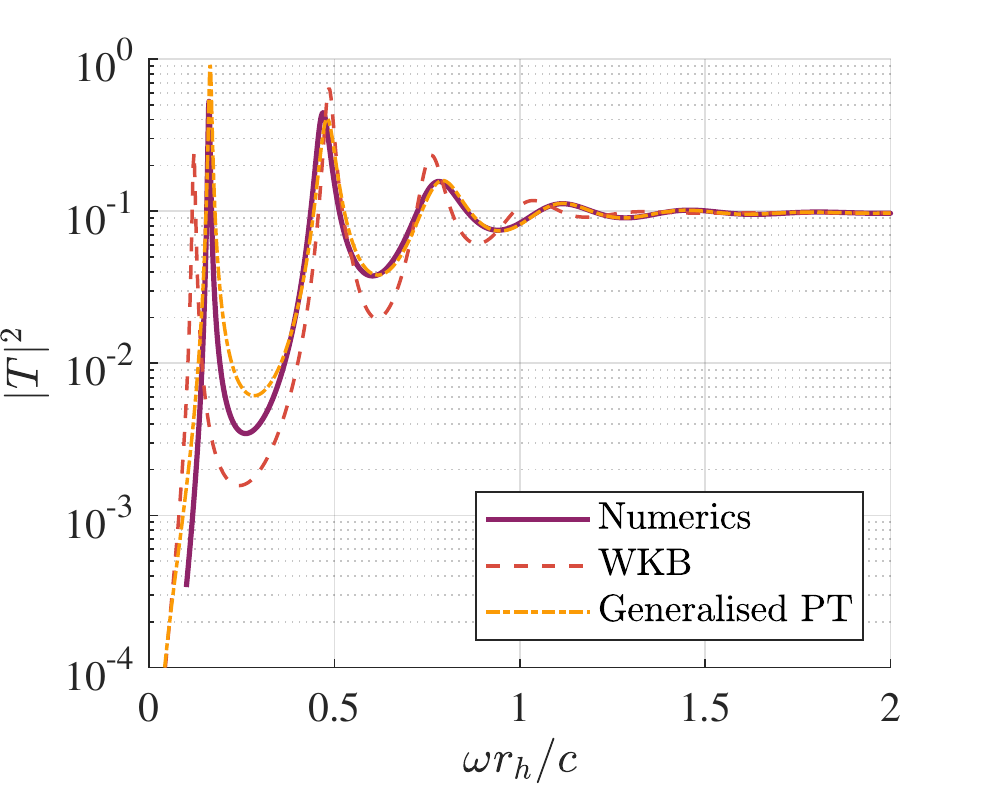}
\caption{Numerical and analytical transmission coefficient for the $m=1$ mode incident on a non-rotating vortex flow ($D=1$) with an inner core located at $r_{*0} = -9$ and with $\mathcal{K}=0.95$. The purple curve depicts the transmission coefficient obtained numerically while the dashed red and dotted dashed yellow curve show the WKB and generalised P\"oschl-Teller approximation respectively. Both analytical predictions agree well with the numerics and capture the essential features.} \label{fig:T_spectrum_comparison}
\end{figure}

\section{Conclusion}\label{sec:conclusion}

In summary, we have seen that the absorption spectrum of a draining vortex with a partially reflective boundary condition in the core exhibits spectral lines. 
The addition of rotation to the system leads to the occurrence of sharp peaks in the reflection coefficient at the bound state frequencies which satisfy the usual superradiance condition $\omega<mC/r_h^2$.
We found that the reflection coefficient had a maximum amplitude for a particular value of $\mathcal{K}$.
The WKB method then revealed that the reflection coefficient was extremised when the amount of superradiant amplification was perfectly balanced by the amount of absorption in the vortex core, in agreement with numerical results.
We also showed how a modified P\"oschl-Teller approximation could be used to obtain analytic expressions for the scattering amplitudes.
Our results show similar features to ECOs in astrophysics \cite{Macedo:2018yoi}.
Similar behaviour has been argued to occur when a Kerr black holes is surrounded by a reflective membrane \cite{maggio2019analytical} and the instability shares similarities with the $w$-mode instability around neutron stars \cite{kokkotas2004w}.

If near horizon reflections occur around real (gravitational) black holes, a consequence would be the occurrence of secondary repetitions of the initial ringdown waveform known as echoes. Although echoes have been the subject of intense research in recent years, e.g.\ \cite{abedi2017echoes,cardoso2019gravitational}, the mechanism underpinning near horizon reflections is usually empirically modelled using the same procedure employed in this paper, that is, by placing a partially reflective boundary condition a small distance from the horizons.
However, in a fluid system, the underlying microscopic theory is known and one could therefore use the fluid to understand the kind of mechanisms that lead to near horizon reflections.
Besides providing insight for gravity, reflections in the vortex core are of intrinsic fluid mechanical interest due to their implications for the stability properties of systems containing vortices e.g.\ \cite{Pu1999,shin2004dynamical,okano2007splitting,isoshima2007spontaneous,giacomelli2020ergoregion,patrick2021quantum,patrick2021origin}.
One approach could be to use spectral lines to learn about the effective field theory inside the vortex core. In particular, we have seen that the separation of the spectral lines is linked, in a first approximation, to the location of the wall in the tortoise coordinates, see Eq.~\eqref{eq:resonance_PT}. The experimental identification of several spectral lines could therefore give a estimate of $r_{*0}$, which in turn can give a precise estimate of the effective inner location in the physical radial coordinate. This is due to the fact that, close to the horizon, for small shifts in $r$ will manifest as large shifts in $r_*$.

Our results may also be able to explain features of the draining vortex experiment in \cite{Torres:2016iee}. 
There, it was observed that the reflection coefficient for the $m=2$ mode was larger than $m=1$ for the flow parameters used.
However, within the simple draining vortex model, the maximum of the reflection coefficient is a decreasing function of $m$, which is true in both dispersive and non-dispersive theories \cite{patrick2020superradiance}.
A potential explanation for this could be that reflections in the vortex core result in peaks in the reflection coefficient, whose locations depend on the value of $m$.
Further evidence for horizon reflections in this experiment was given in \cite{Torres:2020ckk}, where it was shown that the reflecting coefficient for $m<0$ asymptotes to a finite constant at large $\omega$.

\section*{acknowledgments}
TT would like to thank Sam Dolan for useful and inspiring discussions. This work was supported in part by the STFC Quantum Technology Grants ST/T005858/1 (RG \& TT) ST/T006900/1 (SP), STFC Consolidated Grant ST/P000371/1 (RG) and the NSERC (Grant 5-80441 to W.\ Unruh) (SP). RG also acknowledges support from the Perimeter Institute. Research at Perimeter Institute is supported by the Government of Canada through the Department of Innovation, Science and Economic Development Canada and by the Province of Ontario through the Ministry of Research, Innovation and Science.

\bibliography{bibli.bib}
\bibliographystyle{apsrev4-2}

\appendix
\section{Double P\"oschl-Teller potential}\label{app:PT}
Here we aim at modelling the scattering of shallow water waves over a rotating and draining vortex by approximating the potential with a double P\"oschl-Teller potential of the form.
The propagation of an azimuthal mode over a DBT with circulation $C$ and drain $D=1$ is described by the following wave equation,
\begin{equation}
    \partial^2_{r_*} \phi -V(r) \phi = 0,
\end{equation}
with,
\begin{equation}
\begin{split}
    V(r) = & -\left( \omega - \frac{m C}{r^2}\right)^2 \\
    & \qquad + \left(1-\frac{1}{r^2}\right)\left( \frac{m^2 - 1/4}{r^2} + \frac{5}{4r^4}\right).
\end{split}
\end{equation}
The potential $V$ is flat as the tortoise coordinate $r_*$ goes to $\pm \infty$ and has one maximum at $\bar{r}_*$ or in the real radial coordinate $\bar{r}$.
Hence we can split the $r_*$ axis in two regions and define the following potential as a approximation of the true potential,
\begin{equation}
    V_{PT} = \Theta(\bar{r}_* - r_* ) V_1 + \Theta(r_* - \bar{r}_*) V_2,
\end{equation}
where,
\begin{equation}
    V_{1,2}(r_*) = -\left(a_{1,2} + b_{1,2}{\mathrm{sech}}^2\left[ \alpha_{1,2}(r_* - \bar{r}_*). \right]\right).
\end{equation}
This corresponds to a P\"oschl-Teller potential in each region with different parameters $(a_{1,2},b_{1,2},\alpha_{1,2})$.
These parameters are given by,
\begin{eqnarray}
a_1 &=& \left(\omega - m C \right)^2, \\
a_2 &=& \omega^2, \\
b_{1,2} &=& - V(\bar{r_s}) - a_{1,2}.
\end{eqnarray}
The $a$'s are chosen such that the double sech potential has the right asymptotics, the $b$ coefficients are chosen such that the model potential is continuous and the $\alpha_{1,2}$ are chosen such that the potential is differentiable at $\bar{r}_*$, and the curvature matches that of $V$.
Hence the $\alpha$'s are found by solving,
\begin{equation}
    \partial^2_{r_*} V_{1,2}(\bar{r}_*) = \partial^2_{r_*} V(\bar{r}_*).
\end{equation}
This double P\"oschl-Teller approximation to the potential of a draining and rotating vortex is depicted in Fig.~\ref{fig:PT_vs_trueV}

\subsection{Solution to the P\"oschl-Teller potential}
We now want to solve the following equation,
\begin{equation}
    \partial^2_{r_*} \phi -V_{1,2}(r_*) \phi = 0,
\end{equation}
which explicitly is,
\begin{equation}\label{PTr_eq}
    \partial^2_{r_*} \phi +\left(a_{1,2} + b_{1,2}{\mathrm{sech}}^2\left( \alpha_{1,2}(r_* - \bar{r}_*) \right)\right) \phi = 0.
\end{equation}
Introducing a variable $R = \alpha_{1,2}(r_* - \bar{r}_*)$ for each half of the model potential, Eq.~\eqref{PTr_eq} becomes on each side:
\begin{equation}\label{PTR_eq}
    \partial^2_{R} \phi +\frac{1}{\alpha_{1,2}^2}\left(a_{1,2} + b_{1,2}{\mathrm{sech}}^2\left(R\right)\right) \phi = 0.
\end{equation}
The solutions to this equation are given in terms of associated Legendre function, $P^{\mu}_{\nu}$:
\begin{eqnarray}
    f^{in}_{1,2} &=& \Gamma(1-\mu_{1,2}) P_{\nu_{1,2}}^{\mu_{1,2}}(-\text{tanh}(R)), \\
    f^{up}_{1,2} &=& \Gamma(1-\mu_{1,2})P_{\nu_{1,2}}^{\mu_{1,2}}(\text{tanh}(R)),
\end{eqnarray}
where the parameters $\mu_{1,2}$ and $\nu_{1,2}$ are given by,
\begin{eqnarray}
\mu_{1,2} &=& \frac{i\sqrt{a_{1,2}}}{\alpha_{1,2}}, \\
\nu_{1,2} &=& \frac{-\alpha_{1,2} + \sqrt{4b_{1,2}+\alpha_{1,2}^2}}{2\alpha_{1,2}}.
\end{eqnarray}
The ``in'' solution satisfies the ingoing boundary condition at the horizon and the ``up'' solution satisfies outgoing boundary condition at infinity. 
Hence, the general solution is given as a superposition of the ``in" and ``up" solutions,
\begin{equation}
    \phi_{1,2} = A_{1,2} f^\mathrm{in}_{1,2} + B_{1,2} f^\mathrm{up}_{1,2}.
\end{equation}
\subsection{Finding the amplitudes}
At the inner wall, the solution is given by,
\begin{equation}
    \phi_{BC}(r_*\approx r_{*0}) \approx e^{-i\tilde{\omega}r_*} + \left(\mathcal{K}e^{-2i\tilde{\omega}r_{*0}}\right)e^{i\tilde{\omega} r_*}.
\end{equation}
The exact solution in this region is given by,
\begin{equation}
   \phi_1(r_*) = A_{1} f^\mathrm{in}_{1}(r_*) + B_{1} f^\mathrm{up}_{1}(r_*).
\end{equation}
Hence we can find the coefficients $(A_1,B_1)$ by solving the linear system of equations,
\begin{eqnarray}
\phi_1(r_{*0}) &=& \phi_{BC}(r_{*0}), \\
\partial_{r_*}\phi_1(r_{*0}) &=& \partial_{r_*}\phi_{BC}(r_{*0}).
\end{eqnarray}
Then to find the amplitude $(A_2,B_2)$, we match the solution in region I and region II at the top of the potential,
\begin{eqnarray}
\phi_1(\bar{r}_*) &=& \phi_2(\bar{r}_*), \\
\partial_{r_*}\phi_1(\bar{r}_*) &=& \partial_{r_*}\phi_2(\bar{r}_*).
\end{eqnarray}
From the amplitudes $(A_2,B_2)$, we can construct the amplitude of the ingoing and outgoing waves by using the asymptotic expansion of the associated Legendre function,
\begin{equation}
   \Gamma(1-\mu)P^{\mu}_\nu(\text{tanh}(\alpha(r_* - \bar{r}_*))) \sim e^{\alpha\mu r_*} \quad \text{as}\quad r_* \rightarrow \infty,
\end{equation}
as well as the relation between $P^{\mu}_\nu(\rho)$ and $P^{\mu}_\nu(-\rho)$ (see e.g.~\cite{PhysRevD.79.124043}).
Hence we get that,
\begin{eqnarray}
A_\mathrm{out} &=& A_2A_{\mathrm{out},2}^{C=\mathcal{K}=0} + B2, \\
A_\mathrm{in} &=& A_2A_{\mathrm{in},2}^{C=\mathcal{K}=0}.
\end{eqnarray}
Where $A_{\mathrm{out},i}^{C=\mathcal{K}=0}$ and $A_{\mathrm{in},i}^{C=\mathcal{K}=0}$ are respectively the out-going and in-going coefficient in the case of the standard P\"oschl-Teller potential with parameters $(a_i,b_i,\alpha_i)$,
\begin{eqnarray}
A_{\mathrm{out},i}^{C=\mathcal{K}=0} &=& \frac{\Gamma(1- \mu_i)\Gamma(\mu_i)}{\Gamma(-\nu_i)\Gamma(1+\nu_i)} \quad \text{and}\\
A_{\mathrm{in},i}^{C=\mathcal{K}=0} &=& \frac{\Gamma(1- \mu_i)\Gamma(-\mu_i)}{\Gamma(1-\mu_i+\nu_i)\Gamma(-\mu_i-\nu_i)}.
\end{eqnarray}

\subsection{The $u_\infty$ solution}
Using the connection formulae derived above, we can now express the modal coefficients $A_h^{\pm}$ of the $u_\infty$ solution satisfying the boundary condition given in Eq.~\eqref{u_inf} for the generalised P\"oschl-Teller potential.
From the asymptotic at $r_* \rightarrow + \infty$, we have that $(A_\mathrm{in},A_\mathrm{out}) = (0,1)$ which implies that $(A_2,B_2) = (0,1)$.
From the connection at the top of the potential barrier, we get that,
\begin{eqnarray}
    A_1 &=& \frac{1}{2}\frac{\Gamma(1-\mu_2)}{\Gamma(1-\mu_1)}\left(\frac{P_2}{P_1} - \frac{P_2'}{P_1'}\right) \quad \text{and}\\
    B_1 &=& \frac{1}{2}\frac{\Gamma(1-\mu_2)}{\Gamma(1-\mu_1)}\left(\frac{P_2}{P_1} + \frac{P_2'}{P_1'}\right),
\end{eqnarray}
where the coefficients $(P_{1,2},P_{1,2}')$ as the value of the Legendre functions $P_{\nu}^{\mu}(x)$ and its derivative at $x=0$, i.e.
\begin{eqnarray}
    P_{1,2} &=& P_{\nu_{1,2}}^{\mu_{1,2}}(0) \quad \text{and}\\
    P_{1,2}' &=& {P'}_{\nu_{1,2}}^{\mu_{1,2}}(0). 
\end{eqnarray}
Using the asymptotic expansion of the Legendre function as $r_* \rightarrow -\infty$, we can relate the coefficient $(A_1,B_1)$ to $(A_h^{-},A_h^{+})$. Explicitly, we have
\begin{eqnarray}\label{eq:modal_PT_hor}
    A_h^{-} &=& A_1 + B_1 A_{\mathrm{out},1}^{C=\mathcal{K}=0} \quad \text{and} \\
    A_h^+ &=& B_1 A_{\mathrm{in},1}^{C=\mathcal{K}=0}. 
\end{eqnarray}
Note that when $C=0$, we directly have the relations \mbox{$(A_1,B_1) = (A_2,B_2) = (0,1)$}, and we recover the usual modal coefficients of the P\"oschl-Teller model.

\end{document}